# 'Low activation, refractory, high entropy alloys for nuclear applications'

A. Kareer[a], J.C Waite[a], B. Li[a], A. Couet[b], D.E.J. Armstrong[a], A.J. Wilkinson[a]

[a] Department of Materials, University of Oxford, Parks Road, Oxford, OX1 3PH, UK
[b] Department of Engineering Physics, University of Wisconsin Madison, WI, 53706, USA

# Short communication: 'Low activation, refractory, high entropy alloys for nuclear applications'

A. Kareer[a], J.C Waite[a], B. Li[a], A. Couet[b], D.E.J. Armstrong[a], A.J.Wilkinson[a]

[a]Department of Materials, University of Oxford, Parks Road, Oxford, OX1 3PH, UK
[b]Department of Engineering Physics, University of Wisconsin Madison, WI, 53706, USA

**Abstract**

Two new, low activation high entropy alloys (HEAs) TiVZrTa and TiVCrTa are studied for use as in-core, structural nuclear materials for in-core nuclear applications. Low-activation is a desirable property for nuclear reactors, in an attempt to reduce the amount of high level radioactive waste upon decommissioning, and for consideration in fusion applications. The alloy TiVNbTa is used as a starting composition to develop two new HEAs; TiVZrTa and TiVCrTa. The new alloys exhibit comparable indentation hardness and modulus, to the TiVNbTa alloy in the as-cast state. After heavy ion implantation the new alloys show an increased irradiation resistance.

The operating environments envisaged for advanced nuclear reactors create significant challenges for structural materials due to a higher neutron flux, a more corrosive environment and higher operating temperatures [1]. The initial choice material for structural components in Gen IV reactors was 316 austenitic stainless steel, however, due to unacceptable levels of void swelling, the focus has shifted towards reduced activation ferritic/martensitic (F/M) steels, which are suitable for both fission and fusion applications [2, 3]. Although these are promising candidate materials, there remains significant concerns regarding the creep-rupture strength and irradiation embrittlement at 550°C [4]. It is therefore of interest to explore new alloy designs outside the paradigms of conventional steels, to meet these material requirements.

High entropy alloys (HEAs) represent a new class of alloys that have the potential to replace conventional alloys in structural applications. Typically, they consist of four or five alloying elements in close to equiatomic concentrations. The original concept of HEAs was based on the idea that the high configurational entropy of the system would favor the formation of a disordered single-phase solid solution over ordered intermetallic compounds, resulting in simple microstructures with enhanced material properties [5, 6, 7]. Although single phase HEAs have been reported in the literature [8, 9], an even larger number of HEAs have been documented that exhibit complex intermetallic phases [10, 11]. As a result, the concept that configurational entropy would generally stabilise these highly alloyed systems to a single phase is now largely discredited, and research is directed towards forming more ductile HEAs that contain a major solid-solution phase fraction [12, 13]. HEAs based on refractory elements of the 4-5-6 alloy group (named from the groups and periods in the periodic table) show considerable potential for structural applications [14]. Alloys of the NbTaV - (Ti, W, Mo) system, that have a disordered BCC structure, have been proposed as high temperature materials that offer high strength, ductility and oxidation resistance [15, 16]. The alloy TiVNbTa shows excellent compressive mechanical properties at room temperature ($\sigma_y$ = 1273 MPa) and at elevated temperatures ($\sigma_y$ drops to 688 MPa at 900°C) [17]. Variations of the TiVCrZrNb system have also been proposed as high temperature, structural materials which exhibit low densities and high hardness [18]. A concise summary of the current literature on HEAs based on the 4-5-6 elemental palette is tabulated in [19].

HEAs are of interest for nuclear applications due to their claimed 'self-healing' qualities [20, 21]. Experimental irradiation studies using heavy ion irradiation demonstrate that HEAs offer superior radiation resistance compared to conventional alloys, with lower density of dislocation loops and less radiation induced segregation which has been attributed to the severe lattice distortion and sluggish diffusion [22, 23, 24, 25]. More recent work has attributed the irradiation resistance of HEAs to two unique properties; firstly a lower phonon mean free path, limiting the cascade induced heat wave propagation, creating a more localised and longer thermal spike which would favor athermal point defect recombination [26] and secondly, a broadening of the interstitial and vacancy migration energy distributions, with possible overlap of the two distributions, which would favor thermal point defect recombination [27]. Both hypotheses are still under investigation to tentatively explain the potential higher radiation resistance of HEAs. Previous research has demonstrated that the radiation induced volume swelling in the Al$_x$CoCrFeNi alloys is lower than that of conventional nuclear materials. In HEAs, FCC alloys have less swelling than alloys with a BCC + FCC microstructure and single phase BCC microstructures have the largest amount of void swelling. In contrast, conventional alloys with a BCC microstructure would generally have less void swelling than FCC materials [21]. Although the majority of irradiation damage studies have been targeted at FCC HEAs based on transition metals (CoCrFeMnNi), refractory HEAs that offer a low thermal neutron absorption cross-section i.e. TiVZrNb [28] and TiVCrNbMo [29], have been proposed for nuclear applications. However, as pointed out in [19], the suitability of these alloys for use in the next generation of advanced nuclear reactors is somewhat restricted by the high activation elements Nb and Mo. Although activation issues are significantly reduced compared to the most common cobalt containing FCC HEAs, these high activation elements require substantially longer time periods before the radioactivity levels reach a satisfactory limit for 'hands on' maintenance [30]. Materials based on low activation elements, with minimal impurities, have a profound environmental impact by reducing the quantity of high level radioactive waste during decommissioning and allow in-core components to be recycled; low activation is also an essential design criteria for fusion reactors [31, 32]. As such, development of a radiation resistant, low activation HEA, with enhanced high temperature properties would advance the next generation of nuclear reactors.

In this work, we start with the well characterised BCC alloy, TiVNbTa, as this material consists of a single phase, disordered, BCC microstructure, with exceptionally high strength and significant plastic flow in compression at elevated temperatures [17]. The composition is modified by replacing the high activation element, Nb in TiVNbTa with Zr and Cr, to produce TiVZrTa and TiVCrTa; two low activation alloys that have low thermal neutron absorption cross-sections making them suitable for both advanced fission and fusion applications. The nuclear properties of the pure elements are given in Table 1. We report firstly on the microstructure of these alloys and, using nanoindentation, study the effect of heavy ion irradiation on the mechanical properties.

Equiatomic TiVNbTa, TiVZrTa and TiVCrTa alloys were vacuum arc melted in a water cooled, copper crucible using an Arccast Arc200 arcmelter. Raw elements (99.99% purity purchased from Goodfellow, UK) were weighed out to their target stoichiometries to produce a 30g billet, of each com position.

| Element | r (Å) | a (Å) | $T_m$ (°C) | $\sigma_A$ (barns) | Time (years) |
|---|---|---|---|---|---|
| Ti | 1.46 | 3.276 | 1668 | 6.096 | < 3 |
| V  | 1.32 | 3.024 | 1910 | 5.086 | 0.1 |
| Cr | 1.25 | 2.884 | 1907 | 3.05 | < 2 |
| Zr | 1.6  | 3.582 | 1855 | 0.186 | < 3 |
| Nb | 1.43 | 3.301 | 2477 | 1.156 | 30,000 |
| Ta | 1.46 | 3.301 | 3017 | 20.5 | ∼ 10 |

Table 1: Atomic radius (r), Lattice constant (a), melting temperature ($T_m$), thermal neutron absorption cross-section ($\sigma_A$), approximate time (years) for contact dose rate to reach 'hands on' level ($2 \times 10^{-5}$ Sv/h) of the pure Ti, V, Cr, Zr, Nb and Ta elements after 5 years exposure in 3.6GW fusion power reactor [30].

The final microstructure achieved was close to equiatomic (± 2 at.%) measured using EDX. Alloys were melted for approximately 5 minutes and subsequently flipped and remelted to ensure a homogeneous microstructure throughout. Microstructural characterisation of the as-cast material, was carried out using backscatter electron imaging (BSE), energy-dispersive X-ray spectroscopy (EDX) and electron backscatter diffraction (EBSD). A Zeiss Merlin field emission gun scanning electron microscope (FEG-SEM), equipped with an Oxford instruments Xmax 150 EDX detector and a Bruker Quantax EBSD system was used. The crystal structure was identified using X-ray diffraction (XRD) on a Panalytical Empyrean with a 40kV voltage and 40nA current from a Cu K$\alpha$ source.

Heavy ion implantation of a section of each alloy, and a control sample of pure vanadium, was performed at the Surrey National Ion Beam Centre, UK using a 2MeV Tandem accelerator. Vanadium ion implantation was carried out at 500°C using 2MeV V$^+$ ions to a fluence of $2.26 \times 10^{15}$ ions/cm$^2$.
Samples were held at 500°C for 18 hours during the implantation. SRIM (stopping range of ions in matter) software was used to convert the fluence into a displacement per atom (dpa) value. The Ion distribution and quick calculation of damage method was used to obtain a damage profile from 2MeV of vanadium ions in TiVNbTa, yielding a peak damage of 3.6 dpa approximately 700nm below the surface. Nanoindentation was carried out using the continuous stiffness method (CSM) [33] on an Agilent (formerly Keysight, formerly MTS) G200 nanoindenter, fitted with a Berkovich diamond tip. An array of 25 indentations were made to a maximum depth of 1500nm with a strain rate target of 0.05 sec$^{-1}$. The CSM amplitude was 2nm and the frequency 45Hz. Indentation hardness and modulus was measured in both the unirradiated and irradiated portion of each sample to assess the hardening due to irradiation damage and to ensure that the effect of thermal annealing was accounted for.

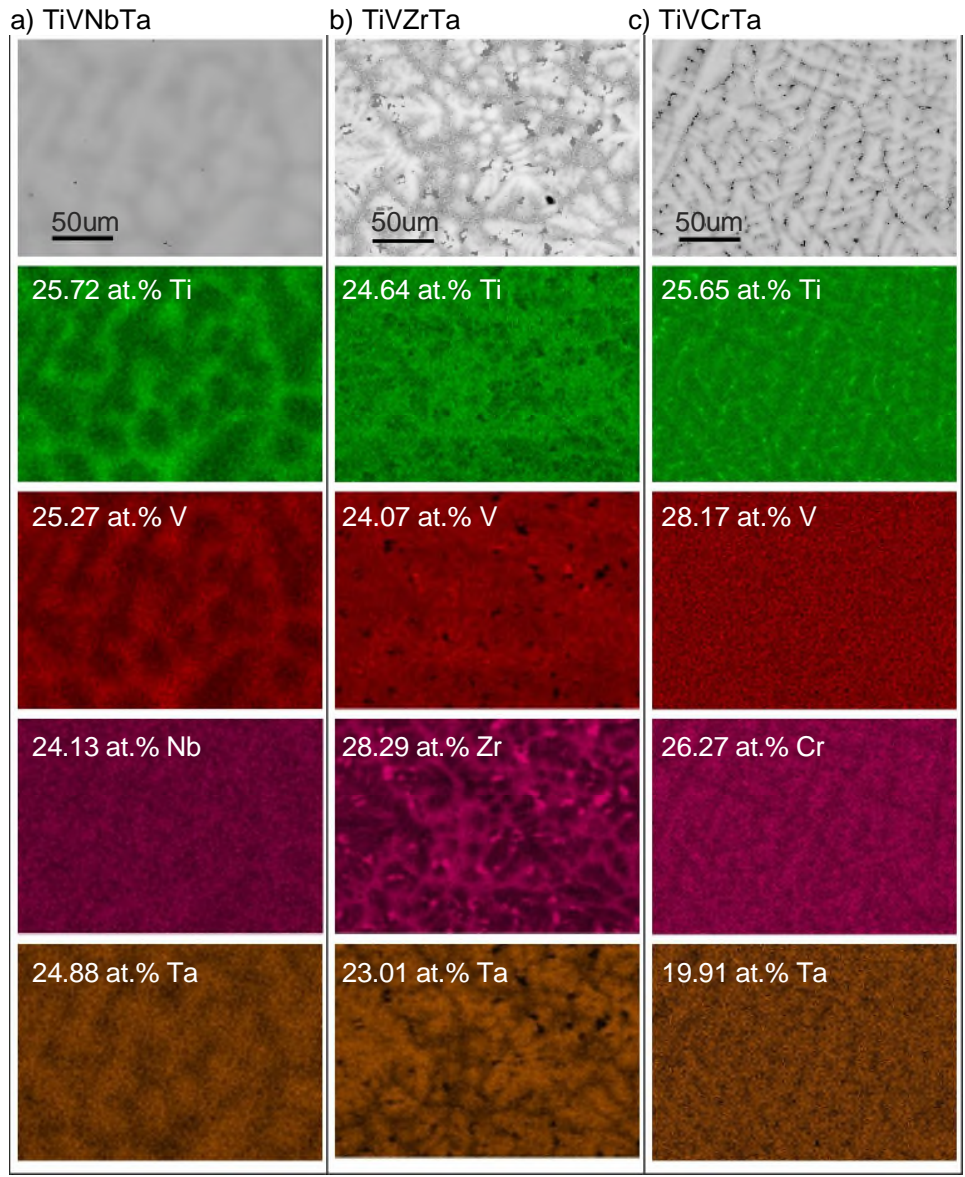

Figure 1: Microstructural analysis of (a) TiVNbTa, (b) TiVZrTa and (c) TiVCrTa showing their respective SEM micrographs (BSE signal) (top) and elemental composition maps, including the nominal composition of each alloy in at.% from EDX (second to fourth rows).

Representative backscatter electron micrographs and EDX composition maps of each alloy are given in Figure 1. The microstructure can vary significantly within an arc melted billet due to lack of a controlled cooling rate, hence in this study, care was taken to ensure that all characterisation was carried out in a consistent location, close to the centre of the 30g billets. The TiVNbTa consists of a typical as-cast dendritic microstructure due to non-equilibrium solidification. Although dendritic structures can be observed in single element metals, it is possible that for these multicomponent alloys, the microstructure is triggered by the different solidification temperatures of each element (see Table 1). EDX maps show qualitatively that the dendrite arms ('brighter' contrast regions in Figure 1 (a)) are enriched with Ta and are depleted in Ti and V; Nb is more uniformly distributed within the alloy matrix, however there is an indication from the EDX maps that the Nb favours the interdendritic, Ti-V rich regions. The TiVZrTa matrix also has a dendritic microstructure of which the arms are enriched in Ta (depleted in Zr, Ti and V). The majority of the interdendritic region is enriched in Zr, Ti and V but depleted in Ta with the exception of a few V rich regions. Uniformly distributed in the matrix are small precipitates which can be identified as the 'dark' spots in the SEM micrograph due to the low average atomic number (see Figure 1(b)). These are enriched in Zr and depleted in V and Ta. The third alloy, TiVCrTa consists of a matrix with a dendritic microstructure. The dendrite arms, again, are enriched in Ta and depleted in Ti and Cr; V is homogeneously distributed in the matrix. A fine distribution of Ti and Cr rich precipitates are observed in the interdendritic region (see dark spots in Figure 1(c)).

Figure 2 shows the indexed XRD spectra for each alloy which indicates all three alloys have a majority, disordered BCC phase. Additional peaks in the EBSD patterns from each region of the microstructures were collected and the band contrast maps are given in Figure 3. The EBSPs from the as-cast TiVNbTa were indexed as a single phase, BCC microstructure with approximate grain size of d=~200 µm and a lattice parameter a = 3.239 Å was obtained from the XRD. In the TiVZrTa, the indexed XRD spectra shows three distinct BCC phases with lattice parameters a = 3.155 Å, a = 3.274 Å and a = 3.470 Å. EBSD is unable to differentiate three individual BCC phases in the microstructure, however it is clear from the band contrast map, Figure 3 (middle), that three distinct regions do exist in this microstructure; that is the matrix region containing the dendritic structure, the precipitates shown as bright white in the band contrast map, due to the high intensity signal obtained from the EBSPs and finally a region where the pattern quality was sufficiently poor within the interdendritic regions. The average grain size of the matrix region is d =~ 50 µm and the precipitates are approximately 7-10 µm in diameter. The TiVCrTa has the largest grain size, d = ~ 600 µm. The matrix is predominantly made up of disordered BCC microstructure (a = 3.11Å from XRD) with a fine distribution of C15, Fd-3m Laves phase precipitates (a= 7.048Å from XRD) forming in the interdendritic regions and along the grain boundaries. These second phase precipitates are approximately 2 µm in size and have been highlighted in red in Figure 3 (right). Table 2 summarises the approximate chemistry (at.%) and the assigned crystallographic structure of the phases present in each alloy system.

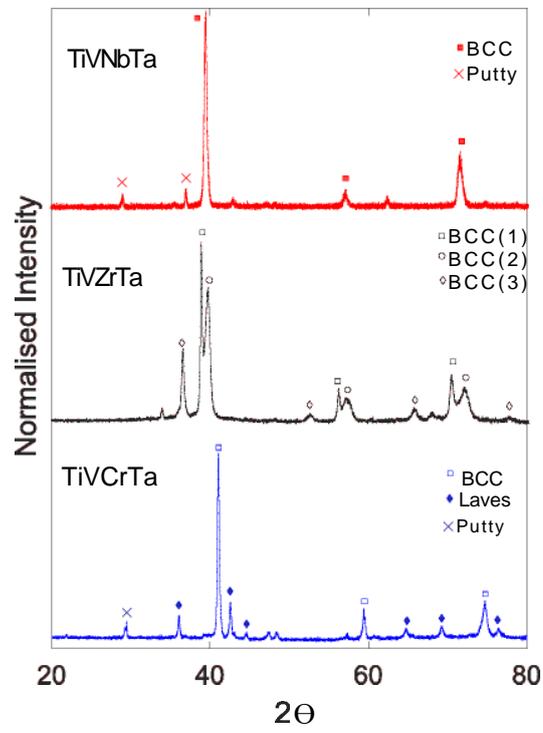

Figure 2: XRD spectra for the three alloys; peaks labelled putty were from the medium used to hold the samples securely in place

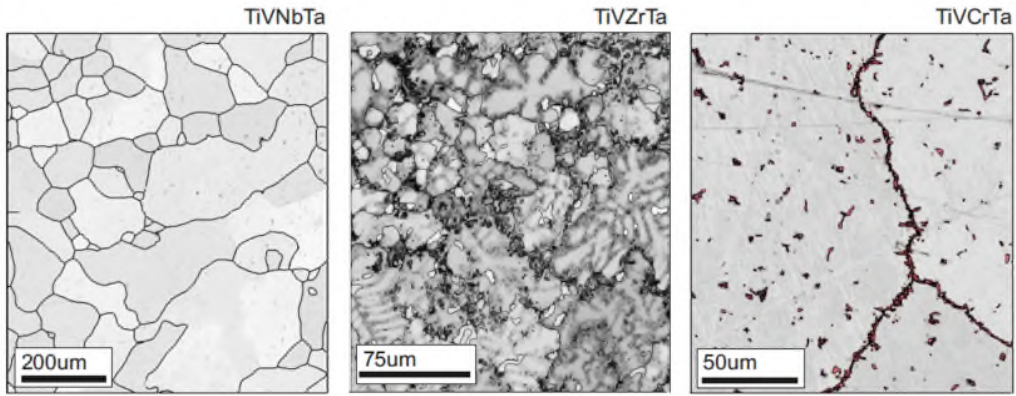

Figure 3: Band contrast maps from EBSD of TiVNbTa (left), TiVZrTa (middle) and TiVCrTa (right). The second indexed phase in the TiVCrTa is highlighted in red

Nanoindentation hardness and modulus as a function of displacement are given in Figure 4(a) and Figure 4(b) respectively for the unirradiated, as-cast HEAs. The TiVZrTa and TiVCrTa have hardness values of 7.41 GPa and 6.83 GPa respectively; both of which are harder than the base alloy TiVNbTa that has a hardness of 5.86 GPa. All three HEAs offer an increased indentation hardness relative to conventional nuclear materials, such as 316SS (2.2 GPa) and T91 (3.1 GPa) [34, 35]. The larger scatter in the indentation data from the TiVZrTa can be attributed to the minor, BCC2, Zr rich phase/precipitates observed for this alloy which had an increased hardness compared to the matrix. Figure 4(c) gives the irradiated hardness data for the HEAs along with the irradiation damage profile, calculated from SRIM for the implantation. The shape of the hardness curve (including an indentation size related increase at shallow depths) is of the same form to that of the unirradiated hardness, indicating no obvious influence from the damaged layer as the plastic zone extends into the unirradiated material beyond the Bragg peak. Note that this damage profile was calculated for the TiVNbTa composition, however, there was negligible differences between the damage profile for the other compositions. The peak damage occurs at a depth of ~700nm below the surface and no damage is expected beyond 1 μm of depth. The irradiation induced hardening, $\Delta H$, is measured at an indentation depth of 300nm, to ensure that the plastically deformed zone is constrained within the damaged layer and thus meaningful values of hardness

| Alloy | Phase/region (Figure 3) | Ti at.% | V at.% | Nb/Zr/Cr at.% | Ta at.% | Structure/lattice parameter |
|---|---|---|---|---|---|---|
| TiVNbTa | Grey (dendritic) | 22.18 (0.76) | 21.88 (0.82) | 25.41 (0.61) | 30.52 (1.73) | BCC (3.239Å) |
|  | Grey (interdendritic) | 30.64 (1.08) | 28.73 (1.22) | 22.42 (0.96) | 18.22 (2.84) | BCC (3.239Å) |
| TiVZrTa | Grey (dendritic) | 24.34 (0.63) | 24.28 (1.14) | 14.66 (2.39) | 36.72 (3.42) | BCC (3.470Å) |
|  | Dark Grey (interdendritic) | 25.48 (0.98) | 23.86 (1.84) | 35.67 (3.59) | 14.99 (2.86) | BCC (3.155Å) |
|  | white precipitates | 17.71 (1.76) | 10.35 (2.83) | 62.36 (13.27) | 9.58 (10.26) | BCC (3.274Å) |
| TiVCrTa | Grey (dendritic) | 23.12 (0.38) | 28.81 (0.19) | 24.54 (0.23) | 23.53 (0.72) | BCC (3.110Å) |
|  | Grey (interdendritic) | 26.60 (1.77) | 27.83 (0.46) | 26.96 (0.92) | 18.61 (2.72) | BCC (3.110Å) |
|  | Red precipitates | 40.74 (7.58) | 22.40 (3.46) | 25.12 (1.72) | 11.74 (6.1) | c15 (7.048Å) |

Table 2: Phases present in each alloy system with respect to the maps shown in Figure 3. Chemical composition of each phase in at.% with standard deviation averaged from 20 EDX spectra in EDX Quant maps (Oxford instuments, AZtec software). Crystal structure and lattice parameter attributed to each phase identified using a combination of EBSD and XRD

are extracted and compared. Additionally, at a depth of 300nm the modulus data is independent of depth indicating that the area function calibration was able to capture the tip geometry well at this depth, with negligible surface and substrate effects (see Figure 4(b)). The hardness of the irradiated and unirradiated material, at an indentation depth of 300nm is plotted in Figure 4(d) along with the ΔH for each alloy. The pure vanadium control sample has an irradiation induced hardening of 1.19 GPa (37%), confirming that the irradiation implantation was carried out successfully. The TiVNbTa shows an irradiation hardening of 0.66 GPa (8%) which suggests some irradiation induced damage accumulation in this single phase alloy, however significantly less than observed in the single element control sample. Irradiation induced hardness changes in the TiVZrTa and TiVCrTa are negligible; a two sample t-test (implemented in Matlab) indicates that in these samples, the two sets of hardness values obtained before and after irradiation come from distributions that cannot be distinguished to the 5% probability level. Hence, no measurable irradiation hardening can be identified.

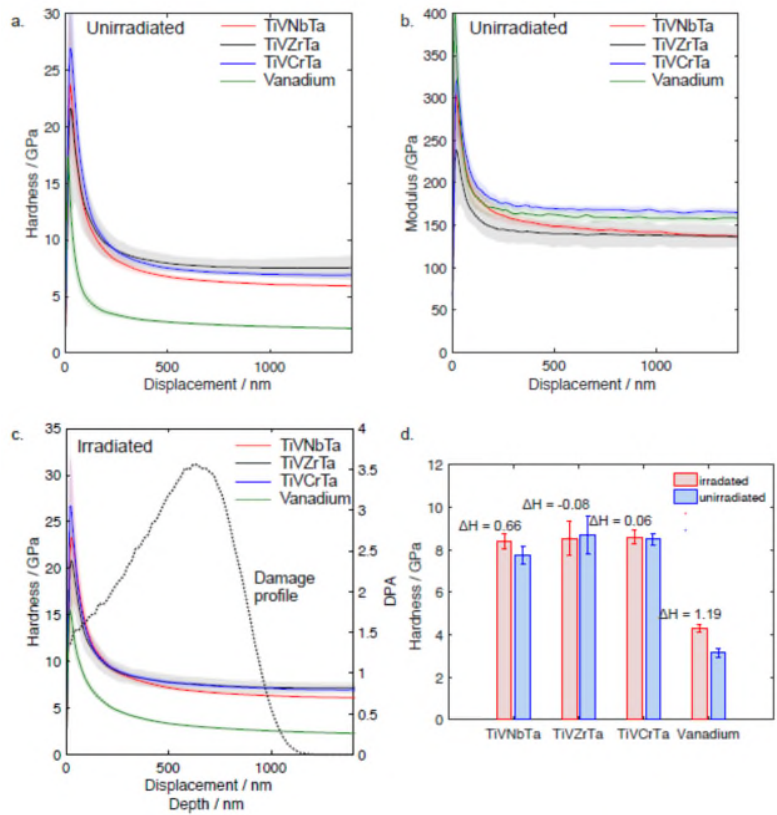

Figure 4: Indentation hardness and irradiation hardening plots (a) unirradiated indentation hardness vs. depth (b) Unirradiated indentation modulus vs. depth (c) Irradiated indentation hardness vs. depth. The irradiation damage profile is given by the dashed line. (d) Indentation hardness of irradiated and unirradiated HEAs and control sample at 300nm indentation depth. Shaded area in (a), (b) and (c) and error bars in (d) represent the standard deviation of 25 indentations.

The new alloys exhibit a majority disordered BCC phase, similar to TiVNbTa, with elemental micro-segregation formed upon solidification due to the different solidification temperatures of the alloying elements (dendritic microstructure). In all three alloys, Ta rich dendrite arms are initially formed due to the high $T_m$. It has been shown for the TiVNbTa alloy that this chemical segregation is alleviated through homogenisation at 1200°C for 72 hours [17]. A theoretical phase prediction was carried out using CALPHAD on the TiVNbTa which has revealed that the microstructure is only single phase down to a temperature of approximately 525°C, after which a second BCC phase starts to precipitate. Multiple BCC phases have been identified theoretically and experimentally for refractory HEAs based on similar compositions [28]. The TiVZrTa has a more complex microstructure and XRD reveals three distinct BCC phases. From the chemical analysis, Zr rich precipitates are identified with a BCC crystal structure which have a significantly stronger EBSD signal, indicating a second minor BCC phase. It is also possible that there exists a V rich third phase in the interdendritic region. The binary phase diagram, Zr-Ta, shows that these elements form two BCC phases at high temperature [36]. It appears the relatively high cooling rates cause these elements to remain in multiple BCC phases, rather than transforming into the low temperature BCC+HCP microsturucture. The crystal structure and lattice parameters assigned to each region in the TiVZrTa microstructure are based on the atomic radius of the element with highest concentration within this region i.e. the BCC phase with the largest lattice parameter has been assigned to the Ta rich phase in the microstructure. TiVCrTa forms an ordered intermetallic phase upon casting that is finely distributed in the major BCC matrix, that are enriched in Ti and Cr. A C15 Laves phase, $Ti_2Cr$, is present in the Ti-Cr binary phase diagram and is favoured over the competing solid solution phase. However, the observed C15 phase in the TiVCrTa consists of (approximately in at.%) 41Ti, 22V, 25Cr, 12Ta, hence some of the Ti could be substituted with Ta and some of the Cr replaced with V. The atomic radius of Cr is small relative to those of the other alloying elements in TiVCrTa which may also favour the formation of the ordered Laves phase in this alloy. More work is needed to assess whether the nucleation and growth of the C15 phase favors formation of fine precipitates that are in thermodynamic equilibrium with the BCC matrix phase.

The measured indentation hardness and modulus of the new alloys are comparable to that of the TiVNbTa, indicating these new alloys will exhibit comparable bulk mechanical properties. The non-homogeneous microstructure of the TiVZrTa and the relative size of the precipitates to the scale of nanoindentation testing, leads to larger experimental scatter in the hardness values; it was identified that the Zr rich BCC2 precipitates have an increased hardness compared to the matrix. As a consequence, this has the potential to reduce the overall ductility of this alloy however, in order to assess this, macro scale mechanical testing is required. The fine dispersion of C15 precipitates observed in the TiVCrTa provide a strengthening mechanism, increasing the hardness relative to the TiVNbTa. It is appropriate to assume that the additional interphases, induced by the second phases, formed in the new alloys act as defect sinks to the irradiation damage, leading to increased irradiation resistance in these alloys.

This work presents three refractory HEAs, two of which consist of elements with low activation and low neutron absorption cross-sections. The microstucture of these alloys consist of a majority disordered BCC phase, with minor secondary phases. The two low activation alloys offer mechanical properties (indentation hardness and modulus) comparable to the well characterised TiVNbTa. However, the new alloys presented in this work offer increased radiation resistance in terms of irradiation induced hardening. Further work, to asses the stability of these phases after homogenisation is needed in addition to a full macro scale mechanical property characterisation.

**Acknowledgements**

This work was supported by the EPSRC funded grant EP/R006245/1; EPSRC Platform grant EP/P001645/1; EPSRC Fellowship EP/R030537/1; and the EPSRC Center for Doctoral Training in the Science and Technology of Fusion Energy.